\documentclass[11pt, a4paper]{article}
 \usepackage{jheparxiv}
\pdfoutput=1
\usepackage{amssymb}
\usepackage{skak}
 \usepackage{dsfont}
\usepackage{float}
\usepackage{nicematrix,tikz}
\usepackage{subcaption}
\usepackage{amsmath}
\usepackage{epsfig}
\usepackage{hyperref}
\usepackage{breakurl}
\usepackage{multirow}
\usepackage{array}
\usepackage{bm}
\usepackage{color}
\usepackage{tikz}
\usepackage[utf8]{inputenc}
\usepackage{braket}
\usepackage{graphicx}
\usepackage{footmisc}
\usepackage{mathtools}
\usepackage{dsfont}

\usepackage{colortbl}
\definecolor{blue2}{cmyk}{1, 0.1, 0.1, 0}

\definecolor{pyBlue}{RGB}{31, 119, 180}
\definecolor{pyRed}{RGB}{214, 39, 40}
\definecolor{pyGreen}{RGB}{44, 160, 44}
\definecolor{pyBlue2}{RGB}{0, 111, 237}
\definecolor{pyRed2}{RGB}{224, 52, 36}

\usepackage{colortbl}
\definecolor{summersky}{cmyk}{0.71,0.33,0,0.5}
\definecolor{flamingo}{cmyk}{0,0.51,0.71,0.5}
\definecolor{rp}{cmyk}{0.2, 1, 0.6, 0}
\definecolor{pacificblue}{cmyk}{0.95,0.3,0, 0.5}
\definecolor{gray60}{cmyk}{0.4,0.4,0,0.8}

\newcommand{\gap}{\phantom{i}}

\renewcommand{\[}{\left[}

\renewcommand{\d}{\mathrm{d}}
\newcommand{\im}{i}
\newcommand{\scri}{\mathcal{I}}
\newcommand{\mrin}{\text{in}}
\newcommand{\mrout}{\text{out}}

\usetikzlibrary{decorations.markings}
\usepackage{cleveref}
\crefformat{section}{\S#2#1#3}
\crefformat{subsection}{\S#2#1#3}
\crefformat{subsubsection}{\S#2#1#3}

\makeatletter
\def\simgt{\mathrel{\lower2.5pt\vbox{\lineskip=0pt\baselineskip=0pt
           \hbox{$>$}\hbox{$\sim$}}}}
\def\simlt{\mathrel{\lower2.5pt\vbox{\lineskip=0pt\baselineskip=0pt
           \hbox{$<$}\hbox{$\sim$}}}}

\makeatother

\def\spa#1.#2{\left\langle#1\,#2\right\rangle}
\def\spb#1.#2{\left[#1\,#2\right]}
\def\sand#1.#2.#3{%
\left\langle#1{\vphantom1}\right|{#2}\left|#3\right]}%
\def\sandmp#1.#2.#3{%
\left\langle#1{\vphantom1}\right|{#2}\left|#3\right]}%
\def\sandpm#1.#2.#3{%
\left[#1{\vphantom1}\right|{#2}\left|#3\right\rangle}%
\def\sandmm#1.#2.#3{%
\left\langle#1{\vphantom1}\right|{#2}\left|#3\right\rangle}%
\def\sandpp#1.#2.#3{%
\left[#1{\vphantom1}\right|{#2}\left|#3\right]}%

\renewcommand{\imath}{\mathrm{i}}

\newcommand{\be}{\begin{equation}}
\newcommand{\ee}{\end{equation}}

\def\S{{\mathbb S}}

\begin{document}

{\baselineskip0pt
\rightline{\baselineskip16pt\rm\vbox to+20pt{
           \hbox{YITP-25-06}
\vss}}%
}

\title{Memory and supertranslations on plane wave spacetimes: an on-shell perspective \\[4pt]\fontsize{10}{0} \selectfont }

\author{Andrea Cristofoli${}^\symqueen$ \&}
\emailAdd{cristofoli@yukawa.kyoto-u.ac.jp}
\author{Sonja Klisch${}^\symknight$}
\emailAdd{s.klisch@ed.ac.uk}
\affiliation{\vspace{0.2cm}${}^\symqueen$Center for Gravitational Physics and Quantum Information,
Yukawa Institute for Theoretical Physics, Kyoto University, 606-8502, Kyoto, Japan \vspace{0.02cm}\\
${}^\symknight$School of Mathematics and Maxwell Institute for Mathematical Sciences,
University of Edinburgh, EH9 3FD, UK }

\abstract{We revisit the computation of the classical gravitational waveform for a particle moving in a plane wave background using on-shell amplitudes. We emphasize the relationship between gravitational memory and the boundary conditions of external scattering states, which were neglected in previous works. We then provide the first tree-level expression for the waveform that captures all memory effects. The waveform is presented in terms of Synge's world function, with explicit tail terms, and a smooth weak memory limit. We also discuss the choice of BMS frame for the waveform on a plane wave background. In flat space, this corresponds to a choice of soft dressing of the initial state. We show that on a plane wave background, this dressing becomes a supertranslation of the waveform, in addition to a phase shift in the waveshape of the background.}

\maketitle

\section{Introduction}
Any solution to the Einstein field equations can be locally approximated as a gravitational plane wave spacetime along a reference null geodesic~\cite{Penrose1976}. Despite being an approximation, gravitational plane waves retain many of the universal features of general relativity, in particular the memory effect ~\cite{Zeldovich:1974gvh, Braginsky:1987kwo}. Due to the Penrose limit they have provided many results valid for any backgrounds such as proof of the Pirani-Bondi theorem on memory effects~\cite{Zhang:2017rno} and the singular behavior of Green's functions along caustics in generic spacetimes~\cite{Harte:2012uw}. Of particular recent interest has been the calculation of observables in general relativity --- such as the gravitational waveform from astrophysical phenomena. On a plane wave spacetime, radiation is already generated by a massive particle following geodesic motion. A modern approach to this calculation is to use the classical limit of on-shell amplitudes on plane waves~\cite{Adamo:2022rmp, Adamo:2022qci, Cristofoli:2022phh}, motivated by corresponding calculations in flat space~\cite{Cristofoli:2021vyo}. 

In this paper, we use on-shell techniques to provide the first detailed treatment of the effects of classical memory on the gravitational waveform of a massive particle passing through a plane wave background. In previous work~\cite{Adamo:2022qci}, the waveform was calculated under the assumption of weak memory where the massive probe follows the same trajectory after passing through the wave. Here, we relax this assumption, considering the more physically realistic, fully non-linear effects of memory. These are captured principally in the Bogoliubov transformations~\cite{Gibbons:1975jb, Garriga:1990dp} of the fields, and require a careful treatment of the in- and outgoing boundary conditions of the problem. The importance of the distinction for calculating observables was recently pointed out in FRW backgrounds~\cite{Aoki:2024bpj}. On gravitational plane waves, the memory effect starkly changes the properties of ingoing and outgoing solutions, see for example~\cite{Harte:2024mwj}. We find that keeping careful track of this behaviour is essential for recovering the correct classical waveform on these backgrounds.

Gravitational waveforms are often ambiguous due to an implicit choice of BMS frame~\cite{Veneziano:2022zwh, Strominger:2013jfa,Boyle:2015nqa}. As an example, on flat space a BMS supertranslation between the intrinsic and canonical BMS frame can be realized on-shell by dressing the initial state with an eikonalised soft charge \cite{Elkhidir:2024izo}. This supertranslation is crucial in comparing the waveform computed in a binary scattering from classical and amplitude based methods~\cite{Georgoudis:2023eke, Georgoudis:2024pdz, Bini:2024rsy}. It is suggestive to study the same dependence on a curved background such as a plane wave background. In this case, we can use a simple relation between the S-matrix on a plane wave background and flat space via a displacement operator~\cite{Cristofoli:2021vyo} to find the equivalent of a BMS supertranslation on a plane wave background. We find that this is expressed as a translation in the retarded Bondi coordinate and by a rotation in the waveshape defining the background. The latter can also be interpreted as a redefinition of the coordinate system of the plane wave. 

The paper is organised as follows: Section \ref{sec:KMOC} provides a brief review of classical and quantum field dynamics on a plane wave backgrounds including Bogoliubov transformations. We propose a representation for fields on such backgrounds at large distances in terms of Bondi coordinates. In Section \ref{sec:waveform}, we calculate the gravitational waveform, with careful consideration of boundary conditions and the memory effect. The result has support on and in the null lightcone parametrised by the Synge's world function (encoding geodesic distance) of the spacetime. In Section \ref{sec:BMS}, we show how ambiguities in the choice of a BMS frame can be implemented by generalizing the results in~\cite{Elkhidir:2024izo} to a plane wave background providing a closed and exact expression for a supertranslated waveform. In Appendix \ref{AppBondi}, we review the asymptotic behaviour of geometric quantities in plane wave spacetime. In Appendix \ref{app:details}, we provide further details of the waveform calculation for interested readers. Finally, in Appendix \ref{app:Synge}, we review Synge's world function on a plane wave background. 

Throughout the paper, we use the mostly negative signature and denote factors of $2\pi$ in integral measures and delta functions by using hats, following the notational and normalization conventions established in~\cite{Kosower:2018adc}
$:
\hat{\delta}^{(n)}(p) :=(2 \pi)^n \delta^{(n)}(p) $ and $ \hat{d}^np:=\frac{d^np}{(2\pi)^n} 
$. On-shell integration measures will be denoted as $d\Phi(p):=\hat{d}^4p \: \hat{\delta}(p^2-m^2)\: \theta(p_0)$.
We work in natural units with $c=\hbar=1$ and the gravitational coupling will be $\kappa:=\sqrt{32 \pi G}$ where $G$ is Newton's constant.

\section{Plane wave backgrounds and Bogoliubov transformations}\label{sec:KMOC}

We start with a brief review of gravitational plane waves in Einstein gravity. For further details on the symmetries and properties of these metrics see~\cite{Blau}. In Kerr-Schild coordinates --- also known as Brinkmann coordinates --- these plane waves have line element
\begin{equation} \label{pwmetric}
\d s^2 = 2 \d x^+ \d x^- - \delta_{ab}  \d x^a \d x^b - H_{ab}(x^-)  x^a x^b (\d x^- )^2 \, ,
\end{equation}
where $a, b = 1, 2$ and $x^{\pm} = (x^0 \pm x^{d-1})/\sqrt{2}$. A plane wave is an exact solution to the vacuum Einstein equations provided that the $ 2\times 2$ matrix $H_{ab}(x^-)$ is trace free: $H_a^a(x^-) = 0$. In order to define a spacetime where asymptotic states are well-defined we will further impose that the wave profile is \emph{sandwich}: there exist initial and final light-front times, $x_i^-$ and $x_f^-$ so that $H_{ab}(x^- < x_i^-) = 0$ and $H_{ab}(x^- > x^-_f) = 0$.

A quantity of interest for us, encoding memory effects on plane waves, are the transverse vielbeins satisfying the differential equation
\begin{equation}
\ddot{E}_{i \, a}  = H_{ab}\, E^{ \gap b}_i, \qquad  \dot{E}^{\gap a}_{ [i}\, E^{\gap b}_{ j]} = 0, \label{vielDef}
\end{equation}
where the Brinkmann indices $a, b$ are raised by the flat transverse metric $\delta^{ab}$. These form a basis of $2$ linearly independent vectors $E_{i \, a}(x^-)$, with index $i = 1, 2$ spanning the set.  The inverse $E^i_{ \gap a}(x^-)$ satisfies
\begin{equation}
E_{\gap a}^{ i} (x^-)\, E_{ i \, b} (x^-) = \delta_{ab},
\end{equation}
with summation implied.
The metric of the geodesic distances is defined using these transverse vielbeins
\begin{equation}
\gamma_{ij}(x^-) \coloneqq E_{(i}^{\gap \, a} (x^-)\, E_{j)\, a} (x^-). \label{ERmetric}
\end{equation}
A further quantity of interest is the deformation tensor encoding the expansion and shear of geodesics on the spacetime via its trace and trace-free parts respectively:
\begin{equation}\label{deformDef}
\sigma_{ab}(x^-) = \dot{E}^{i}_{\gap a} E_{i \, b}.
\end{equation}

\medskip

 The function $H_{ab}(x^-)$ uniquely defines the plane wave, however all other quantities defined above feature some coordinate ambiguity. More precisely, \eqref{vielDef} gives a set of unique solutions provided we have appropriate initial conditions. Natural ones to consider are the \emph{in-going} and \emph{out-going} initial conditions~\cite{Gibbons:1975jb, Garriga:1990dp}
\begin{equation}
E^{\text{in}}_{i \, a} (x^- < x_i^-) = \delta_{i \, a}, \qquad E^{ \text{out}}_{i \, a} (x^- > x_f^-) = \delta_{i\,  a}. \label{vielAsymps}
\end{equation}
These have the general opposite asymptotics
\begin{gather}
E^{\text{in}}_{i \, a} (x^- > x_f^-) = b^{ \text{in}}_{i\, a} + x^- \, c^{ \text{in}}_{i\, a}, \label{oppAsympsIn} \\ E^{ \text{out}}_{i\, a} (x^- < x_i^-) = b^{ \text{out}}_{i\,\gap a} + x^- \, c^{ \text{out}}_{i\,\gap a}  \, ,\label{oppAsymps}
\end{gather}
which satisfy
\begin{equation}
b_{[i}^{a \, \text{in}}\, c_{j] \, a}^{\text{in}} = b_{[i}^{a \, \text{out}}\, c_{j] \, a}^{\text{out}} = 0, \qquad c_i^{\gap a \, \text{out}} = \delta_{ib}\, \delta^{bj} \,  c_j^{\gap c \, \text{in}}.
\end{equation}
This is imposed from the second equation in \eqref{vielDef} and the conservation of the Wronskian between the two solutions. Additionally, rotating the coordinates, $c_{i \, a}$ can be set to be a symmetric matrix. The matrix $c$ is the key quantity encoding the memory effect in this paper --- in fact it encodes the gravitational velocity memory effect~\cite{Bieri:2024ios} as seen in~\cite{Adamo:2022rmp} from amplitude calculations. Zero memory can be imposed by taking all entries $b_{i\, a}\rightarrow \delta_{i\, a}, c_{i \, a} \rightarrow 0$, which was the assumption in~\cite{Adamo:2022qci}. In the following we will stay to all orders in the memory. Notice that the zeroes of the determinant of \eqref{oppAsympsIn} correspond to the focusing of initially parallel geodesics in the out-region. These $x^-$-hyperplanes are the caustics of solutions to the wave equation considered in the next section. Following~\cite{Garriga:1990dp} we consider only wave profiles $H_{ab}$ such that the zeroes of $\det(E^{\mrin/\mrout})$ occur only outside of the wave region $[x_i^-, x_f^-]$. With this assumption, it can be shown that $\det(E^{\mrin})$ has at most two zeroes (counting multiplicity). We will relate the number of zeroes $m$ and $\sqrt{\det(c)}$ by specifying the branch of the square root
\begin{equation}
\sqrt{\det(c)} = e^{i m \pi/2} \sqrt{|\det(c)|}.
\end{equation}
This follows from the form of $\sqrt{\det(E^\mrin)}$ at large $x^-$ with appropriate analytic continuation through its zeroes.

\subsection{Classical fields on plane wave backgrounds}

We are interested in solutions to the wave equation on plane wave spacetimes that have initial conditions
\begin{equation}
\phi^{\mrin}_{p} (x^- < x^-_i) = e^{\im p \cdot x}, \quad \phi^{\mrout}_{p} (x^- > x^-_f) = e^{\im p \cdot x}.
\end{equation}
These solutions are given explicitly~\cite{Ward:1987ws} by
\begin{multline}\label{eq:classical-sol}
    \phi_{p}^{\mrin/\mrout}(x) = \frac{1}{\sqrt{|E^{\mrin/\mrout}|}} \, \exp \Big[ \im \Big( p_+ x^+ + \frac{p_+}{2} \sigma_{ab}^{\mrin/\mrout} + p_i E^{i \, \mrin/\mrout}_a x^a + \frac{p_i p_j}{2 \, p_+} F^{ij}_{\mrin/\mrout} \\+ \frac{m^2}{2p_+} x^-\Big)\Big]
\end{multline}
where $|E^{\mrin / \mrout}| = \det E^{\mrin / \mrout}$, with geometric quantities as defined in (\ref{vielAsymps},~\ref{deformDef}) with the appropriate boundary conditions as in \eqref{oppAsymps}, and 
\begin{align}
    F^{ij}_{\mrin}(x^-)& \coloneqq \int^{x^-}_{x_i^-} \gamma^{ij}(s) \d s  - \delta^{ij} x_i^-, \\
    F^{ij}_{\mrout}(x^-) &\coloneqq \int^{x^-}_{x_f^-} \gamma^{ij}(s) \d s + \delta^{ij}x_f^-.
\end{align}
Solutions with higher spin (in particular gravitational radiation) can be constructed out of the scalar solution by applying a spin-raising operator~\cite{MasonSpin, Adamo:2017nia, Araneda:2022lgu}
\begin{equation}\label{eq:dressed-pol}
\!\!\mathcal{E}^{\eta}_{\mu \nu}(k;x)\, dx^{\mu}dx^{\nu} 
=  \bigg[\bigg(\epsilon^a\!\left(\frac{k_j}{k_+} E_a^j+\sigma_{a b} x^b\right) \!\mathrm{d} x^-
+\epsilon_a \mathrm{d} x^a \bigg)^2 
-\frac{\mathrm{i}}{k_+} \epsilon_a \epsilon_b \sigma^{a b}\, (\mathrm{d} x^-)^2\bigg] \phi_{k}(x),
\end{equation}
where labels for in/out are left implicit.

\medskip

\begin{figure}
\centering
\includegraphics[width=0.48\textwidth]{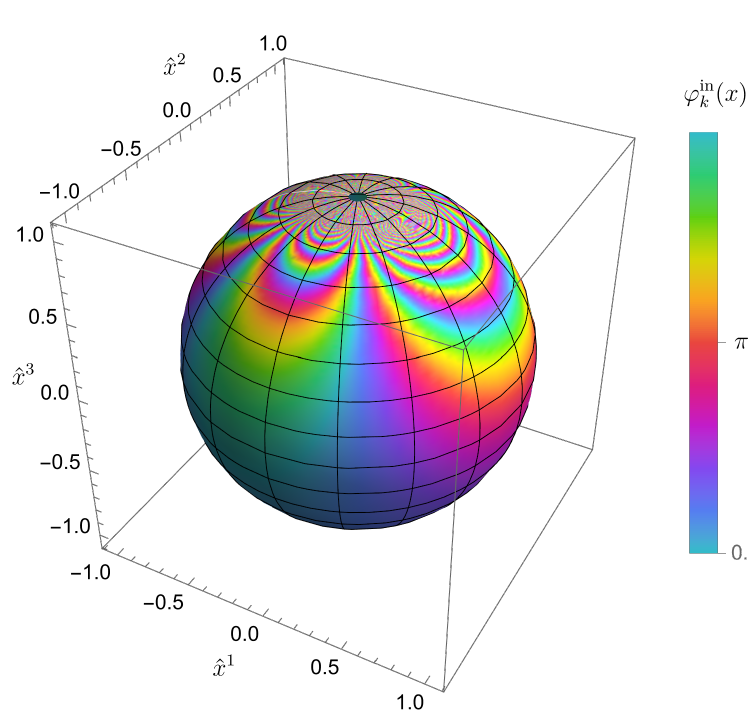}
\caption{The behaviour of the phase $\varphi_{k}^{\mrin}(x)$ defined in \eqref{eq:asyphase} on the future celestial sphere at $u = 0.6$, with parameters $k_1 = 0.5$, $k_2 = - 0.125$, $k_+ = 1$. The north pole corresponds to $\hat{x}^3 = 1$. The behaviour of this phase should be contrasted to the everywhere rapidly phase of the typical $e^{i k \cdot x}$, though this is compensated by the overall $1/r$ behaviour of the solution \eqref{asympScalarBehav}}.\label{fig:wavefronts}
\end{figure}

It will be important in subsequent calculations that the in-going and out-going solutions for generic memory on a plane wave background have completely different asymptotic behaviour. Here we will focus on massless fields. For similar considerations, see~\cite{Harte:2024mwj}. For our application, this can be concretely analysed by transforming the solutions from Brinkmann $(x^+, x^-, x^a)$ to retarded Bondi coordinates $(u, r, \hat{x})$
\begin{equation}
 x^+ =\frac{u + r(1 + \hat{x}^3)}{\sqrt{2}}, \quad x^- =\frac{ u + r (1 - \hat{x}^3)}{\sqrt{2}}, \label{BrinkToBondi}
\end{equation}
and $x^a = r \hat{x}^a$. Details on this transformation applied to plane waves can be found in Appendix \ref{AppBondi}. We find that the leading large-$r$ behaviour of the ingoing scalar wavefunction  of massless momentum $k$ in the asymptotic future is given by
\begin{equation}
     \phi^{\mrin}_{k}(x) \sim \frac{\sqrt{2}}{\sqrt{\det (c)}\, r\, ( 1- \hat{x}^3)}  e^{\im \varphi_{k}^{\mrin}(x) }\label{asympScalarBehav}
 \end{equation}
 where the phase is defined up to an $\hat{x}$-independent constant as
 \begin{equation}
 \varphi_{k}^{\mrin}(x) \coloneqq  \,k_+ \Big(\frac{ \sqrt{2} u}{1 - \hat{x}^3} + \frac{ \hat{x} \cdot (c^{-1})b \cdot \hat{x}}{(1 - \hat{x}^3)^2} \Big)  + \frac{  \,\sqrt{2} \,  k_i (c^{-1})^i_{\, a} \hat{x}^a }{1 - \hat{x}^3}.\label{eq:asyphase}
 \end{equation}
 The overall suppression by $r$ (and no rapidly oscillating-in-$r$ phase) is due to the $x^- \rightarrow \infty$ behaviour of the vielbeins $E^{\mrin}$ in \eqref{oppAsympsIn}. We stress that for  a generic plane wave, the $\sqrt{\det (c)}$ appearing in \eqref{asympScalarBehav} can be real or imaginary depending on the number of caustics. A visual representation of the wavefronts of the phase \eqref{eq:asyphase} for an example initial momentum can be seen in Figure \ref{fig:wavefronts}.  Gravitational states built using spin-raising operators have the same properties,  and we will make use of this to avoid applications of the saddle-point approximation\footnote{Though it should be noted that this approximation also neglects terms when $k \sim 1/r$, due to the nature of the large-$r$ expansion in \eqref{eq:asyphase}. This is similar to the issues mentioned recently in \cite{Elkhidir:2024izo, Jorstad:2024yzm}
In Section \ref{sec:BMS} we investigate how to account for these soft contributions in a similar way to~\cite{Elkhidir:2024izo}.} in our waveform calculation in Section \ref{sec:waveform}.

\subsection{Quantized fields on plane wave backgrounds}
Given a proper Hilbert space $\mathcal{H}$, a quantized field operator $\hat{\mathbf{\Phi}}(x): \mathcal{H} \rightarrow \mathcal{H}$, in momentum space coordinates adapted to lightfront ones, is described by a general mode expansion as
\begin{equation}
\begin{aligned}
\hat{\mathbf{\Phi}}(x) = \int \frac{\hat{d} p_{+} \, \hat{d}^2 p_{\perp} }{2p_{+}} \: 
\Big( \hat{a}_{p} \phi_{p}(x) 
+ \hat{a}_{p}^{\dagger} \phi_{p}^{*}(x) \Big)  ,
\end{aligned}
\end{equation}
where the modes $\phi_{p}(x)$ satisfy the classical equations of motion on a plane wave background as \eqref{eq:classical-sol} while the annihilation and creation operators obey the standard commutation relations 
\begin{equation}
\left[\hat{a}_{p}, \hat{a}_{l}^{\dagger}\right]=2p_{+}\hat{\delta}\left(p_{+}-l_{+}\right) \hat{\delta}^{(2)}(p_{\perp}-l_{\perp}),    
\end{equation}
with all other commutators being zero. One of the subtleties of QFT on a non-trivial background generally is that there is not a unique set of natural mode functions for expanding the scalar field, or equivalently, there is no preferred set of creation and annihilation operators that span $\mathcal{H}$. This is at the heart of phenomena such as pair production in black hole spacetimes. In the case of plane wave spacetimes, we can choose mode functions $\{\phi_{p}^{\text{in}}\}$ which behaves as plane waves in the incoming region $x^- <x_i^-$; alternatively, another natural set of mode functions previously described is $\{\phi_{p}^{\text{out}}\}$ which behave as plane waves in the outgoing region $x^- > x_f^-$. The two choices define two inequivalent sets of creation and annihilation operators $\{\hat{a}^{in}_{p},(\hat{a}^{in}_{p})^{\dag}\}$ and $\{\hat{a}^{out}_{p},(\hat{a}^{out}_{p})^{\dag}\}$. The relation between the two can be found by noticing that both sets of modes are a basis of solutions to the equations of motion. In particular, since the positive and
negative frequencies are not mixed \cite{Gibbons:1975jb, Garriga:1990dp} we have the following relation between in/out modes as well as ``in" and ``out" operators using Bogoliubov coefficients defined as
\begin{equation}\label{eq:scalar-2nd-quan}
\phi_{k}^{\mathrm{in}}(x)=\int \frac{\hat{d} l_{+} \hat{d}^2 l_{\perp}}{2l_{+}} \alpha_{k_{+}, l_{+}}(k_{\perp},l_{\perp}) \phi_{l}^{\mathrm{out}} (x) \, ,
\end{equation}
\begin{equation}\label{eq:Bogo-a}
\hat{a}_{k}^{\mathrm{out}}=\int \frac{\hat{d} l_{+} \hat{d}^2 l_{\perp}}{2l_{+}} \alpha_{k_{+}, l_{+}}(k_{\perp},l_{\perp}) \hat{a}_{l}^{\mathrm{in}} .
\end{equation}
Non-trivial Bogoliubov coefficients generally imply the existence of two inequivalent notions of vacuum, defined as the state annihilated respectively by $\hat{a}_{k}^{\mathrm{in}}$ or $\hat{a}_{l}^{\mathrm{out}}$. However, since there is no pair production in this theory, the two vacua are the same: $\ket{0_{out}}=\ket{0_{in}}$.

\medskip

The Bogoliubov coefficients can be easily obtained by comparing the solutions for the in and out modes along a constant $x^- = x_0^- > x_f^-$ slice of the spacetime \cite{Garriga:1990dp,Adamo:2017nia}
\begin{equation}\label{alpha-plus}
\alpha_{k_{+}, l_{+}}(k_{\perp},l_{\perp})=2 l_{+}\hat{\delta}\left(k_{+}-l_{+}\right) \frac{\mathrm{e}^{-\mathrm{i}\left(s_l+r_{k, l}\right)}}{\sqrt{\det (c)}} \, ,
\end{equation}
where 
\begin{gather}
s_{k, l} \coloneqq  - \frac{k_i k_j}{2k_+} F^{ij}_{\mrin}(x_0) + \frac{l_i l_j x_0^-}{2l_+}, \\
r_{k, l} \coloneqq \frac{1}{2k_+} (l_a - k_i E^{i\, \mrin}_a)(\sigma^{-1}_{\mrin})^{ab} (l_b - k_j E^{j\, \mrin}_b)|_{x_0}.
\end{gather}
The coefficients \eqref{alpha-plus} seem dependent on the slice $x_0$, but it can be shown that they are invariant under changes in choice of time slice.

\medskip 

\section{The exact waveform on a plane wave background} \label{sec:waveform}

\begin{figure}
\centering
\includegraphics[width=0.53\textwidth]{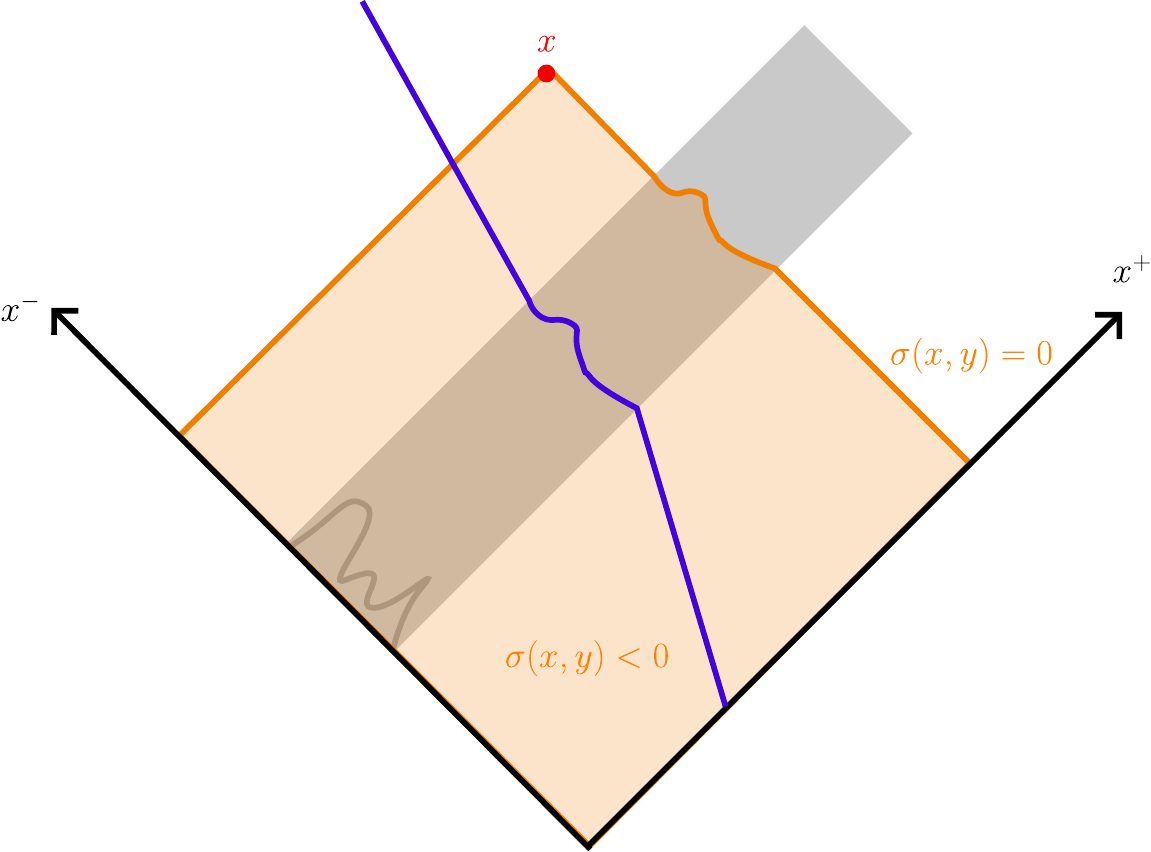}
\caption{An idealised depiction of the tail effect on a gravitational plane wave with $m =0$, projected onto the $(x^+, x^-)$ plane. The solid orange boundary represents points where Synge's world function $\sigma(x, y) = 0$ and is the past lightcone of the point $x$. On the other hand, the shaded orange region has $\sigma(x, y)<0$. Green's functions for gravitational radiation in plane waves have support in both regions~\cite{Harte:2013dba}, violating Huygen's principle. Support from the $\sigma(x, y)>0$ region is often known as the `tail'. The blue line represents an example massive geodesic.}\label{fig:tail}
\end{figure}

Having now introduced the major elements of classical and quantum fields on plane wave backgrounds, we proceed with the computation of the gravitational waveform emitted by a point particle undergoing geodesic motion on such backgrounds using on-shell techniques. One important difference with respect to \cite{Adamo:2022qci} is that we will not apply any weak memory approximation. First of all, to describe a free point particle in the in-region, we introduce the following semi-classical state
\begin{equation}\label{eq:free-state}
    \ket{\Psi}=\int d\Phi(p) e^{i p \cdot b}\phi(p) \ket{p_{in}}, \quad \ket{p_{in}}:=\hat{a}^{\dag,in}_{p}\ket{0_{in}}.
\end{equation}
The quantity $\phi(p)$ is a wavepacket sharply localized on a given on-shell momentum $p_{0}$ as in \cite{Kosower:2018adc}. The four vector $b_{\mu}$ instead denotes a relative distance of the particle with respect to the chosen coordinate system defining our metric \eqref{pwmetric}. The evolution of (\ref{eq:free-state}) is unambiguously described in the non-trivial outgoing region with memory in terms of the $\mathcal{S}$-matrix on a plane wave background
\begin{equation}\label{eq:S-dress}
    \mathcal{S}_{\alpha} \ket{\Psi}= \int d\Phi(p) \, e^{i p \cdot b} \, \phi(p) \, \mathcal{S}_{\alpha} \ket{p_{in}} , \quad
    \mathcal{S}_{\alpha}= \hat{\mathbb{C}}_{\alpha}^{\dag} S \hat{\mathbb{C}}_{\alpha}  .
\end{equation}
We have also introduced a unitary displacement operator to highlight the relation between perturbation theory on this background and the usual standard flat spacetime approach, denoted by the $S$-matrix on flat spacetime
\begin{equation}
    \hat{\mathbb{C}}_{\alpha} := \exp \left[-\frac{1}{2} \int d \Phi(k) |\alpha_{\eta}(k)|^2\right]
    \times \exp \left[ \int d \Phi(k) \alpha_{\eta}(k) \hat{a}_{k,\eta}^{in,\dagger}\right] \,.
\end{equation}
Here the sum over the helicities $\eta$ is implied. In computing the classical waveform we first need to define the graviton operator that is being measured at $\scri^+$. Similarly to \eqref{eq:scalar-2nd-quan}, we can write this operator in two equivalent ways
\begin{equation}\label{eq:in/out-wave}
\begin{aligned}
    \mathbf{\hat{h}}_{\mu \nu}(x) &= \int d\Phi(k) \sum_{\eta = 1}^2 \mathcal{E}^{in \, \eta}_{\mu \nu}(k;x) \hat{a}^{in}_{k,\eta} + h.c \\
    &= \int d\Phi(k) \sum_{\eta = 1}^2 \mathcal{E}^{out \, \eta}_{\mu \nu}(k;x) \hat{a}^{out}_{k, \eta} + h.c. \, .
\end{aligned}
\end{equation}
Here $\mathcal{E}^{in \, \eta}_{\mu \nu}(k;x)$ and $\mathcal{E}^{out \, \eta}_{\mu \nu}(k;x)$ are solutions to the linearized Einstein field equations on a plane wave background with incoming and outgoing boundary conditions, respectively. In what follows, we choose incoming boundary conditions. With our choice, the large 
$r$ behavior of (\ref{eq:in/out-wave}), keeping
$u$ fixed, can be inferred from the limits discussed in the first section of (\ref{eq:dressed-pol}). Indeed, the leading-order contribution to the waveform (evaluated in the outgoing region) is
\begin{multline}\label{eq:gravop}
    \mathbf{\hat{h}}_{\mu \nu}(x)\Big|_{\scri^+} 
    = \frac{\sqrt{2}}{r \, (1 - \hat{x}^3)} \Re \bigg( 
    \int \frac{\hat{d} k_+ \hat{d}^2 k_{\perp}}{k_+ \sqrt{\det (c)}} 
    \sum_{r = 1}^2 \epsilon^{a\,r} \epsilon^{b\,r} 
    \Big[\frac{ \sqrt{2}\, \hat{x}_a n_{\mu}}{1 - \hat{x}^3} + \delta_{a \mu} \Big] 
    \Big[\frac{\sqrt{2}\, \hat{x}_b n_{\nu}}{1 - \hat{x}^3} + \delta_{b \nu} \Big]  \\ \times 
    \hat{a}_{k, r}^{in} e^{i \varphi_k^{in}(x)} 
    \bigg)\Big|_{\scri^+} \, ,
\end{multline}
where, recall \eqref{eq:asyphase}, $\varphi^{in}_k(x)|_{\scri^+}$ has no scaling in $r$ in the out-region, and $c$ is the $2 \times 2$ matrix capturing the behaviour of the vielbein \eqref{oppAsymps} and therefore the memory effect in this background. This also means that the `tail' is subleading in $1/r$ in the expansion of the gravitational perturbations, while the polarisations are no longer dependent on $k$ (which would have before assured Lorentz gauge). The operator now satisfies Lorentz gauge to leading order in $r$. Most importantly, let's notice that the definition of the waveform operator itself on $\scri^+$ carries memory contributions due to $\phi^{in}_k(x)|_{\scri^+}$ and $c$. As for the classical waveform, and as emphasized in \cite{Aoki:2024bpj}, we should then expect not only contributions from on-shell amplitudes but also from the definition of the waveform operator itself 
\begin{equation}\label{eq:waveform-FRW-amplitudes-def}
    h_{\mu \nu}(x)|_{\mathcal{I}^{+}}:=\lim_{\hbar \rightarrow 0} \bra{\Psi}\mathcal{S}^{\dag}_{\alpha}\hat{\mathbf{h}}_{\mu \nu}(x)\mathcal{S}_{\alpha}\ket{\Psi}\Big|_{\scri^+} \ .
\end{equation}
The radiation emitted by a particle along geodesic motions is described by a $3$-point amplitude. Thus, this observable is defined by the following building block
\begin{equation}
\label{eq:eq:wave-a-adag}
 \bra{\Psi}\mathcal{S}^{\dag}_{\alpha}\hat{a}^{in}_{k,\eta}\mathcal{S}_{\alpha}\ket{\Psi} 
    = \int \hat{d}^4q \: \hat{\delta}(2 q \cdot p_0) 
    \times \bra{p_{0,in} + q_{in}, k^{\eta}_{in}} \mathcal{S}_{\alpha} \ket{p_{0,in}} \, ,
\end{equation}
where the absence of a $2$-point is related to having introduced a completeness relation spanned by massive particles with incoming boundary conditions. We can then use results for such amplitudes from \cite{Adamo:2020qru} in order to describe the exact asymptotic waveform as a frequency integral over a Gaussian function in $k_{\perp}$ 
\begin{equation}
    h_{\mu \nu}(x)\Big|_{\scri^+} =  
    \frac{ \sqrt{2} \, \kappa}{r \, (1 - \hat{x}^3)p_{+}}\, \, \Re  \bigg( \frac{e^{-i m \pi/2}}{\sqrt{|\det(c)|}} \int_{\mathbb{R}} dy^{-} \int_{\mathbb{R}^+} \frac{\hat{d} k_+}{k_+} \: I_{\mu \nu}(x,y^{-}, k_+) 
    \bigg) \, ,\label{eqhForm}
\end{equation}
The function $I_{k_{+}}(u,y^{-})$ is defined in the Appendix \ref{app:details} and has two components with distinct scalings in $k_+$:
\begin{equation}
I_{\mu \nu} (u, y^-, k_+) = \frac{2 \pi }{\sqrt{\det A}} \Big(k_+ N_{1\, \mu \nu} + N_{2\, \mu \nu}\Big) e^{i k_+ \bar{\sigma}(x, y^-)},
\end{equation}
with coefficients $N_{1, 2}$ and matrix $A$ defined in Appendix \ref{app:details}, and are themselves independent of $k_+$.  Here, we report its integral over the frequency, as this is the main object of interest for us. The integral depends on the value of $m$ --- the number of caustics in the out-region, defined in Section \ref{sec:KMOC}. This will dictate the functional structure as a consequence of taking the real part in \eqref{eqhForm} according to the four-fold pattern~\cite{Harte:2012uw}.
The integral depends on  \emph{Synge's world function} $\sigma(x, y)$ through the reduced quantity $\bar{\sigma}(x, y) \coloneqq - \sigma(x, y)/(x^- - y^-)$. Synge's world function measures half of the geodesic distance squared between two points on the spacetime manifold~\cite{Synge:1960ueh, Poisson:2011nh}. We specifically consider $\sigma(x, y^-)$  to measure the geodesic distance between a point $(u, \hat{x})$ on $\scri^{+}$, and a particle at light-front time $y^-$, following geodesic motion with initial momentum $p$. This encodes important properties of the geometry of plane wave background and has been extensively studied in the study of caustics and wave propagation on these backgrounds \cite{Harte:2013dba, Harte:2012uw}. Most importantly, it dictates the support of Greens functions on the background. We define the reduced Synge's world function of interest in \eqref{eq:appRedSynge}. Defining $h_{\mu \nu}^{m}(x)$ as the waveform from a background with $m$ caustics, for $m=0$ or $m=2$ we have
\begin{multline}
    h^{m}_{\mu \nu}(x)\Big|_{\scri^+} = (-1)^{m/2}
    \frac{ \sqrt{2} \pi \, \kappa}{r \, (1 - \hat{x}^3)p_{+}\sqrt{|\det (cA)|}} \\
 \times  \int_{\mathbb{R}} \!\d y^{-} \big(N_{1\, \mu \nu} \delta[\bar{\sigma}(x,y^{-})] + N_{2\, \mu \nu} \Theta[\bar{\sigma}(x,y^{-})]\big).
\end{multline}
 From this expression we recover the tail effects prominent in~\cite{Adamo:2022qci} - violations of Huygen's principle for gravitational radiation on plane wave backgrounds~\cite{Harte:2013dba} - in the appearance of $\Theta[\bar{\sigma}(x, y^-)]$. Concretely, this represents the contributions to the classical radiative observable from the interior the past null lightcone of the observer, rather than only the boundary. See Figure \ref{fig:tail}. 

The single remaining integral has a part that is localised on solutions to $\bar{\sigma}(x, y^-) = 0$ which can be determined analytically in specific examples such as impulsive plane waves, and numerically in generic cases. The part containing $\Theta[\bar{\sigma}(x, y^-)]$ takes the form of an integral over the light-front coordinate $y^-$ governing the motion of the particle.
The main differences between the result of~\cite{Adamo:2022qci} and of the current paper is that the presence of memory is apparent through the expression via the matrix $c$ appearing in the definitions of $N_1, N_2, A,\bar{\sigma}$ in (\ref{eq:memN1}, \ref{eq:memN2}, \ref{eq:barA}, \ref{eq:appRedSynge}). Its presence changes the functional form of the waveform through Synge's world-function, and terms emerge that would not have been present otherwise in the integrand, highlighted in Appendix \ref{app:details}. We have therefore recovered the first all-orders in gravitational memory result for a scattering waveform on an exact gravitational plane wave.

Instead, for $m =1$ we have 
\begin{multline}
    h^{m}_{\mu \nu}(x)\Big|_{\scri^+} = 
    \frac{ \sqrt{2} \pi \, \kappa}{r \, (1 - \hat{x}^3)p_{+}\sqrt{|\det (cA)|}} 
     \\
     \times  \int_{\mathbb{R}} \!\d y^{-} \big(N_{1\, \mu \nu} \mathsf{pv} (\bar{\sigma}(x,y^-)^{-1}) + N_{2\, \mu \nu} \mathsf{pv}(\log \bar{\sigma}(x,y^{-}))\big),
\end{multline}
The transformation taking $\delta[\bar{\sigma}(x, y^-)] \rightarrow \mathsf{pv}(\bar{\sigma}(x, y^-)^{-1})$ and $\Theta[\bar{\sigma}(x, y^-)] \rightarrow \mathsf{pv}(\log \bar{\sigma}(x, y^-))$ when considering a background with an additional caustic hypersurface matches the results for the four-fold structure of Green's functions on plane waves found in~\cite{Harte:2012uw, Cho:2023dnf} (see \cite{Casals:2009zh,Dolan:2011fh} for examples on more generic spacetimes with caustics).

\section{The choice of BMS frame on a plane wave background} \label{sec:BMS}

We conclude by examining how the choice of a BMS frame influences our previous calculation.  As noted in \cite{Elkhidir:2024izo}, one choice of BMS frame can be incorporated within the amplitude framework by dressing the initial state with an eikonal resummation of on-shell 3-point amplitudes, corresponding to a linearised Schwarzschild field for the ingoing particles at large distances. In flat spacetime, this procedure results in a supertranslation of the waveform. To understand the implications of this soft dressing in a plane wave background, let us first briefly review the relation between supertranslations and on-shell amplitudes, as derived in \cite{Elkhidir:2024izo}. First, consider the dressing of the initial state generated by the exponentiation of the following soft charge
\begin{equation}
\begin{aligned}
Q_{S}:=\sum_\eta  \int \mathrm{d} &\Phi(k) \hat{\delta}(2 p \cdot k) \mathcal{A}_3\left(p, k_\eta\right) a_\eta^{\dagger}(k) + \text { h.c. }  \: ,
\end{aligned}
\end{equation}
where $\mathcal{A}_3$ describes an on-shell 3-point amplitude for a massive particle emitting a graviton on a flat spacetime. The initial state can now be considered as
\begin{equation}\label{eq:dressing}
\ket{\Psi_{D}}:=e^{i Q_{S}}\ket{\Psi} \: .
\end{equation}
The impact of the dressing on the observable such as the waveform, can be seen by introducing the hard charge \cite{Elkhidir:2024izo}
\begin{equation}\label{eq:hard-charge}
\begin{aligned}
Q_H &=   \sum_\eta \int \mathrm{d} \Phi(k) H(k) a_\eta^{\dagger}(k) a_\eta(k) .  \\
H(k) & := -2G p \cdot k  {\left[\log \left(\frac{p \cdot k}{m k^0}\right)+Y_{\varepsilon}\right]} \: .
\end{aligned}
\end{equation}
This can also be viewed as a weighted occupation number operator\footnote{The quantity $Y_{\epsilon}$ is a function of a dimensional regulator and it is divergent in four dimensions. It can be easily removed by a choice of origin of the retarded time when computing the waveform.}.
Using these charges, the authors of~\cite{Elkhidir:2024izo} saw that the soft dressing induces a shift in the retarded time by a function of the angles on $\scri_{+}$ --- the Veneziano-Vilkovisky supertranslation~\cite{Veneziano:2022zwh}. 

\medskip

The same formalism can also be applied to waveform calculations on plane wave backgrounds. The key observation is that the S-matrix on a plane wave background can be related to the S-matrix on a flat space time by a displacement operator which generates a coherent state as in (\ref{eq:S-dress}).
We can now reconsider the calculation of the waveform from the previous section, adding a soft dressing to the initial state as in \cite{Elkhidir:2024izo}. Since both approaches deal with an asymptotically flat region at early times, the dressing will be the same and given by (\ref{eq:dressing}). The mean value that defines the waveform is then
\begin{equation}
\begin{aligned}\label{eq:soft-wave}
h_{\mu \nu}(x)&=\bra{\Psi_{D}}\mathcal{S}_{\alpha}^{\dag} \mathbf{\hat{h}}_{\mu \nu}(x)\mathcal{S}_{\alpha}\ket{\Psi_{D}} \\
& =\bra{\Psi}e^{-i Q_S}\mathcal{S}_{\alpha}^{\dag} \mathbf{\hat{h}}_{\mu \nu}(x)\mathcal{S}_{\alpha} e^{i Q_S}\ket{\Psi} \: .
\end{aligned}
\end{equation}
We now introduce the same hard charge as in flat space (\ref{eq:hard-charge}) which annihilate the initial state. Using this, we can rewrite the waveform only in terms of quantities defined on a flat spacetime 
\begin{equation}
\begin{aligned}\label{eq:wave-plane}
h_{\mu \nu}(x)=\bra{\Psi}e^{-i Q_T}\mathcal{S}_{\alpha}^{\dag}\mathbf{\hat{h}}_{\mu \nu}(x)\mathcal{S}_{\alpha} e^{iQ_{T}}\ket{\Psi} \: ,
\end{aligned}
\end{equation}
where
\begin{equation}\label{eq:pro-cohe}
Q_{T}:=Q_{S}+Q_{H}\,, \quad [S,Q_T]=0 \,, \quad [Q_S,Q_H]=0 \ .
\end{equation}
In a flat spacetime $\alpha=0$ and we could have then easily moved the total charge closer to the graviton operator using \eqref{eq:pro-cohe} obtaining one of the main results in \cite{Elkhidir:2024izo}
\begin{equation}\label{eq:flat}
e^{-i Q_T} \hat{a}_\eta(k) e^{i Q_T}=e^{i \omega T(n)} \hat{a}_\eta(k)+ \delta(2 p \cdot k) \mathcal{A}_3 \hat{\mathds{1}} \ ,
\end{equation}
where $T(n)=2 G p \cdot n(\log (p \cdot n/m)+Y_{{\varepsilon}})$ and $n^{\mu}=k^{\mu}/\omega$. The presence of the supertranslation in the observable is then reflected in how the annihilation and creation operators transform under the action of the exponentiated total charge: the soft charge generates the potential mode while the hard charge generate a supertranslation of the measured waveform on $\scri_{+}$. 

However, when a background field is present as a coherent states the above analysis is modified. First, note that it's possible to rewrite (\ref{eq:soft-wave}) only in terms of quantity defined on a flat spacetime as follows
\begin{equation}
\begin{aligned}
h_{\mu \nu}(x)=\bra{\Psi,\alpha}e^{-iQ_S}S^{\dag}
\mathbf{\hat{h}}_{\mu \nu}(x)Se^{iQ_S}\ket{\alpha, \Psi}+h^{\alpha}_{\mu \nu}(x) \ .
\end{aligned}
\end{equation}
where we have used $\hat{\mathbb{C}}_\alpha^{\dag} \hat{\mathbf{h}}_{\mu \nu} (x)\hat{\mathbb{C}}_{\alpha} = \hat{\mathbf{h}}_{\mu \nu}(x) + h^{\alpha}_{\mu \nu} (x) \hat{\mathds{1}}$  and $h^{\alpha}_{\mu \nu}$ is the background metric. We can now insert the identity as $e^{iQ_H}e^{-iQ_H}$ between the initial state and the soft charge operator, using the identities \eqref{eq:pro-cohe} to find
\begin{equation}
\begin{aligned}
h_{\mu \nu}(x)=\bra{\Psi,\alpha}e^{iQ_H}e^{-iQ_T}S^{\dag}
 \quad \quad \quad \quad \quad \quad \\ \mathbf{\hat{h}}_{\mu \nu}(x)Se^{iQ_T}e^{-iQ_H}\ket{\alpha, \Psi}+h^{\alpha}_{\mu \nu}(x) \ .
\end{aligned}
\end{equation}
Using the fact that $[Q_T, S] = 0$ we may move the total charge next to the waveform operator, making use of the property (shown in~\cite{Elkhidir:2024izo}) that
\begin{equation}
e^{-i Q_T } \hat{\mathbf{h}}_{\mu \nu}(x) e^{i Q_T} = \hat{\mathbf{h}}_{\mu \nu}(x) |_{u + f(z, \bar{z})} + h_{\mu \nu}^{\text{Schw}}(x) \hat{\mathds{1}}
\end{equation}
where $f(z, \bar{z})$ describes the Veneziano-Vilkovisky supertranslation of the waveform and $h_{\mu \nu}^{\text{Schw}}(x)$ is the linearised Schwarzschild metric of the initial particle.

Up to now, the waveform when we insert a soft dressing on the initial state is
\begin{equation}
h_{\mu \nu}(x)=\bra{\Psi,\alpha}e^{iQ_H}S^{\dag}
\mathbf{\hat{h}}_{\mu \nu}(x)Se^{-iQ_H}\ket{\alpha, \Psi}_{u+f(z,\bar{z})}+h^{\alpha}_{\mu \nu}(x)+h^{\mathrm{Schw}}_{\mu \nu}(x) \ .
\end{equation}
The main difference now with respect to the case on a flat background is the presence of the hard charge $e^{-i Q_H}$ acting on the coherent state of the background. Interestingly, it can be shown using properties of the hard charge \eqref{eq:hard-charge} --- viewed as a weighted occupation number operator --- and the coherent state, that 
\begin{equation}
e^{-i Q_{H}} \ket{\alpha} =   \ket{\alpha e^{-iH}} \ .
\end{equation}
This holds for a generic coherent state, but inserting the plane wave form~\cite{Cristofoli:2022phh} $\alpha(k) = \delta^2 (k_{\perp}) g(k_-)$ and localising onto the $\delta$-function this shifts the frequency profile to
\begin{gather}
g(k_-) \exp [ -i k_- A (p_+)],\label{eq:hard-phase} \\
A(p_+) = - 2G p_+ \Big[  \log \Big( \frac{\sqrt{2} p_+}{m}\Big) + Y_{\varepsilon} \Big].
\end{gather}
The quantity $A(p_+)$ is real and therefore this corresponds in position space to a large coordinate transformation of the $x^-$ coordinate of the plane wave background. In particular, this is the same as the Veneziano-Vilkovisky supertranslation at $\hat{x}^3 = 1$ but now also extended into the bulk. 
\medskip 

The result is that the waveform from dressing the in-going particle can be expressed as
\begin{equation}\label{eq:main-result}
h_{\mu \nu}(x)=\bra{\Psi,\alpha e^{-i H}}S^{\dag}
\mathbf{\hat{h}}_{\mu \nu}(x)S\ket{\alpha e^{-i H}, \Psi}_{u+f(z,\bar{z})}+h^{\alpha}_{\mu \nu}(x)+h^{\mathrm{Schw}}_{\mu \nu}(x) \ .
\end{equation}
In summary, the soft dressing on a plane wave background induces a Veneziano-Vilkovisky supertranslation  of the original waveform as well as a rotation in the coherent waveshape defining the plane wave background. Note that much of this analysis did not require restricting to a plane wave background and would follow similarly for any other gravitational background that can be represented as a coherent state. Plane wave backgrounds however allow us to have a concrete interpretation of the hard phase \eqref{eq:hard-phase} as a coordinate transformation of the $x^+$ coordinate.

\section{Conclusions}

We have revisited the computation of the classical waveform from massive scattering on plane wave backgrounds using on-shell amplitudes. This has taken the form of two aspects: considering plane waves beyond the `miraculous'~\cite{Zhang:2024uyp} zero-memory case considered in~\cite{Adamo:2022qci}; and the effects of a soft dressing of the initial state in the amplitudes calculation, following~\cite{Elkhidir:2024izo}. In addressing the former, we have seen how a key object in gravitational optics --- Synge's world function measuring geodesic distance --- emerges naturally from amplitudes calculations. In addition, our result reemphasises the importance of the tail effect~\cite{Harte:2012uw} in gravitational physics, and showcases explicit dependence on the displacement and velocity memory effect. Further, considering the second aspect, we saw that dressing the initial state generates not only a BMS supertranslation of the waveform (as demonstrated in flat space in~\cite{Elkhidir:2024izo}) but also a simple coordinate shift of the background metric. 

Gravitational plane waves are a particularly suitable and fruitful playground for exact calculations of observables in general relativity. Their strength is that --- despite their simple structure --- they demonstrate many key aspects of generic backgrounds such as memory (which can be interpreted generally as non-trivial geodesic motion on the spacetimes), and tail contribution to their Green's functions. We hope that  having a careful understanding of key observables on plane wave backgrounds from the on-shell perspective of amplitudes will pave a way towards calculations on more generic spacetimes.

\section*{Acknowledgements}
We are grateful to Tim Adamo, Katsuki Aoki, Abraham Harte and Donal O'Connell for useful conversations. SK was supported by an EPSRC studentship and the Simons Collaboration on Celestial Holography, while the work of A.C. was supported by JSPS KAKENHI Grant No.~JP24KF0153.

\appendix 
\section{Brinkmann and Bondi coordinates} \label{AppBondi}
   In this appendix we review the coordinate transformation from Brinkmann to Bondi coordinates, and the asymptotic behaviour of the geometric quantities at play on a plane wave background at large distances.
 
For an asymptotically flat spacetime, Bondi coordinates are chosen so that the metric components have suitable large $r$ behaviour. Different coordinate choices that satisfy this condition are related to each other via BMS transformations (supertranslations and superrotations). In our case we are not in an everywhere asymptotically flat spacetime, as can be seen by the Ricci scalar for a plane wave. However, we may define a preferred set of coordinates for \emph{asymptotically plane wave} spacetimes which is flat asymptotically, in the sense of Bondi, in all directions except for a set of measure zero at the boundary. \emph{Ab initio} the metric is therefore \eqref{pwmetric}, where we now relate the coordinates to Bondi coordinates in the usual way for lightfront coordinates. 
 
 \medskip
 
Bondi coordinates may be defined from Cartesian coordinates in four dimensions via
\begin{gather}
u = t - r, \quad v = t + r,\\ r = \sqrt{(x^1)^2 + (x^2)^2 + (x^3)^2}, \quad \hat{x}^i = \frac{x^i}{r}
\end{gather}
where $\hat{x}^i$ parametrise the celestial sphere $S^2$. The \emph{flat} metric in terms of the retarded coordinates $(u, r, \mathbf{\hat{x}})$ is
\begin{equation}
\d s^2 = (\d u)^2 + 2 \,\d u \,\d r - r^2 \,\d \Omega_{2}
\end{equation}
whilst the flat metric parametrised by the advanced coordinates $(v, r, \mathbf{\hat{x}})$ is
\begin{equation}
\d s^2 = (\d v)^2 -2 \, \d v \,\d r - r^2 \, \d \Omega_{2}
\end{equation}
where $\d \Omega_2 = \d \phi^2 + \sin^2 \phi \,  \d \theta^2$ is the flat metric on $S^2$. It will sometimes be useful for us to use spherical polar coordinates $(\phi, \theta)$ where $(\hat{x}^1, \hat{x}^2, \hat{x}^3) = (\cos \theta \sin \phi, \sin \theta \sin \phi, \cos \phi)$. There are two antipodally related points that will be important for us: where $\hat{x}^3 = 1$ (the north pole) and where $\hat{x}^3 = -1$ (the south pole).

The transformation from lightfront coordinates to Bondi coordinates is then done using
\begin{gather}
 x^+ =\frac{u + r(1 + \hat{x}^3)}{\sqrt{2}}, \quad x^- =\frac{ u + r (1 - \hat{x}^3)}{\sqrt{2}}, \label{advCoordTransf}\\
 x^+ = \frac{v - r(1 - \hat{x}^3)}{\sqrt{2}}, \quad  x^- =\frac{ v - r (1 + \hat{x}^3)}{\sqrt{2}}, \label{retCoordTransf}
\end{gather}
with the transverse coordinates transforming trivially. The plane wave metric in bulk in the adapted Bondi gauge is thus
\begin{multline}
\d s^2 = (\d u)^2 + 2\, \d u \, \d r - r^2 \d \Omega_2 
+ r^2 H_{ab}\big[ u + r (1 - \cos \phi)\big] \\ \times\hat{x}^a \hat{x}^b \Big( \d u + (1 - \cos \phi)\,  \d r +  r \sin \phi\,  \d \phi \Big)^2. \label{advBondi}
\end{multline} 
in terms of retarded coordinates, and
\begin{multline}
\d s^2 = (\d v)^2 - 2\, \d v \, \d r - r^2 \d \Omega_2 
+ r^2 H_{ab}\big[v - r (1 + \cos \phi)\big]\\ \times \hat{x}^a \hat{x}^b \Big( \d v + (1 + \cos \phi)\,  \d r +  r \sin \phi\,  \d \phi \Big)^2. \label{redBondi}
\end{multline}
 in terms of advanced coordinates. The second line in each of the above equations captures the curved aspect of these spacetimes. It is now instructive to look at the $r \rightarrow \infty$ limit, keeping $u$ ($v$) constant. Focussing on \eqref{advBondi}, when $\phi \neq 0$, the $r \rightarrow \infty$ limit takes the argument of $H_{ab}$ to $\infty$. Due to the sandwich condition on the metric, the curved term vanishes and we recover flat space as expected. However, when $\phi = 0$ the coordinates are both not well-defined, and we run into ambiguities taking $r \rightarrow \infty$ whilst also taking $\phi \rightarrow 0$. In contrast to the usual flat metric, this singularity in the curved metric is incurable, and we content ourselves with it being undefined in this limit. In fact, this is just the statement that the metric is not asymptotically flat in the $\phi = 0$ null direction.

 Having identified how to take large-$r$ limit in these background, we collect in Table \ref{asympTab} the asymptotic behaviours of the geometric quantities appearing in the calculations at $\scri^\pm$. Of particular interest is the dependence of these quantities on the velocity memory matrix $c$ associated with the background. In particular, the limit of taking $c \rightarrow 0$ and $r \rightarrow \infty$ do not commute for many of these quantities so we need to treat the memory effect carefully.  
 \begin{table}[t]
\centering
\begin{tabular}{l | l}
\multicolumn{1}{c|}{$\scri^-$} & \multicolumn{1}{c}{$\scri^+$}  \\ \hline
$\displaystyle E_{i \,a} =\delta_{i \, a} $ & $\displaystyle E_{i \, a} = \frac{c_{i \, a} ( 1- \hat{x}^3) r}{\sqrt{2}} + b_{i \, a} + \frac{u}{\sqrt{2}} c_{i \, a} + \ldots$ \\
$\displaystyle E^{i}_{\gap a} = \delta^i_a$ & $\displaystyle  E^{i}_{\gap a} = \frac{\sqrt{2}(c^{-1})^i_{\gap a}}{r (1 - \hat{x}^3)} + \ldots$\\
$\displaystyle \gamma_{ij} = \delta_{ij} $ & $\displaystyle \gamma_{ij} =  \frac{c_{i a} c^{a}_{\gap j} r^2 (1 - \hat{x}^3)^2}{2} + \ldots$\\
$\displaystyle \gamma^{ij} = \delta^{ij} $ & $\displaystyle \gamma^{ij} =  \frac{2 (c^{-1})^i_{\gap a} (c^{-1})^{j \, a}}{r^2 (1 - \hat{x}^3)^2} + \ldots$\\
$\displaystyle \sigma_{ab} = 0$ & $\displaystyle \sigma_{ab} = - \frac{\sqrt{2} \delta_{ab}}{r ( 1- \hat{x}^3)} + \frac{2 (c^{-1})^i_{\gap a}b_{i\, b} + \sqrt{2}\delta_{ab}}{r^2(1 - \hat{x}^3)^2} + \ldots $ \\
$\displaystyle F^{ij} = \delta^{ij}\, x^- $ & $\displaystyle F^{ij} = \text{const. }+ \ldots$\\
$\displaystyle  \det E_{i \, a} = 1$ & $ \displaystyle  \det E_{i \, a}= \frac{1}{2}\det (c)\,  r^2 ( 1- \hat{x}^3 )^2 + \ldots$ \\
\end{tabular}
\caption{Asymptotic large-$r$ behaviour of relevant quantities at $\scri^{\pm}$ for incoming boundary conditions \eqref{oppAsympsIn}}\label{asympTab}
\end{table}

\section{Details of the waveform calculation} \label{app:details}

In this appendix we provide a more detailed calculation of the gravitational waveform, coming from the on-shell momentum integral of a 3-point amplitude and a graviton operator:
    \begin{equation}
    \label{appheq}
    h_{\mu \nu}(x)\Big|_{\scri^+} = \frac{ \sqrt{2}\kappa}{r \, (1 - \hat{x}^3)p_{+}} 
    \times \Re \frac{1}{\sqrt{\det(c)}}\int_{x^-_i}^{x^-_f} dy^{-}\int_{\mathbb{R}^+}\frac{\hat{d} k_+}{k_+} \: I_{\mu \nu}(x, y^{-},k_{+}).
\end{equation}
The $y^-$ integral here is over the finite interval $[x^-_i, x^-_f]$ of the wave, as this is the only region where the scalar feels a force and emits radiation\footnote{Restricting to this zone corresponds to neglecting boundary terms from the flat regions.}. Here we have hidden the $k_{\perp}$ integral in the definition of the term
\begin{multline} \label{eq:Idef}
     I_{\mu \nu}(x, y^{-},k_{+}):=- i\sum_{\eta = 1}^2 \epsilon_a^{\gap\eta} \epsilon_b^{\gap\eta} \Big[\frac{\sqrt{2}\,\hat{x}_a n_{\mu} }{1 - \hat{x}^3}+ \delta_{a \mu} \Big]\Big[ \frac{\sqrt{2} \,\hat{x}_b n_{\nu}}{1 - \hat{x}^3} + \delta_{b \nu} \Big] \\ \times  \int d^{2}k_{\perp} P_{\alpha}(y) P_{\beta}(y) \, \mathbb{P}^{\alpha \beta cd}_{\mrin}(k,y^{-}) \epsilon^{- \eta}_{c}\epsilon^{-\eta}_{d} \times  e^{- \im \mathcal{V}_{k, p} (y^-) + \im  \varphi_k^{\mrin}(x) - \im k \cdot b}\Big|_{\scri^+}.
\end{multline}
The last two lines here are proportional to the 3-point amplitude on a plane wave background. This involves the dressed momentum of particles in this spacetime, defined for generic masses and with in-going quantities as~\cite{Adamo:2017nia, Adamo:2020qru}
\begin{multline}
P_{\alpha}(y) \d y^{\alpha}\coloneqq p_+ \d y^+ + (p_i E^{i}_a + p_+ \sigma_{ab} y^b) \d y^a \\
+ \Bigg( \frac{m^2}{2p_+} + \gamma^{ij} \frac{p_i p_j}{2p_+} + \frac{p_+}{2} \dot{\sigma}_{bc} y^b y^c + l_i \dot{E}^i_b y^b \Bigg)\d y^-
\end{multline}
and the polarization dressing for a graviton of momentum $k$  where
\begin{gather}
\mathbb{P}_{\mrin}^{\alpha \beta cd} \coloneqq \mathbb{P}^{\alpha c} \mathbb{P}^{\beta d} -i n^{\alpha} n^{\beta} \sigma^{cd}/k_+, \\
\mathbb{P}^{\mu \nu} \coloneqq g^{\mu \nu}(y) - 2 K^{(\mu}(y) n^{\nu)}/k_+. 
\end{gather}
Importantly, roman letters only span the \emph{transverse} directions. This in practise means that the $P_-(y)$ component of the dressed momentum never enters the calculation.  In addition, the Volkov exponent is (to leading order in $\hbar$)
\begin{equation}
\mathcal{V}_{k, p} (y^-) = \frac{y^- m^2 k_+}{2 p_+^2} +\Bigg[ \frac{p_i p_j k_+}{2p_+^2} - \frac{p_j k_i}{p_+} + \frac{k_i k_j}{2k_+}\Bigg] \times  \int^{y^-} \d s \,  \gamma^{ij}(s) 
\end{equation}
whereas the phase $\varphi_k^{\mrin}(x)$ is defined in \eqref{eq:asyphase}, and $b$ is the impact parameter of the scalar with respect to our coordinate system. 
To simplify the expression in \eqref{eq:Idef} we use the properties of the transverse polarisation sum for gravity 
\begin{equation}
\sum_{\eta = 1}^2 \epsilon^{\gap \eta}_a \epsilon^{\gap \eta}_b \epsilon_{c}^{- \eta} \epsilon_{d}^{- \eta} = \delta_{ac}^{\perp} \delta_{b d}^{\perp} - \frac{1}{2} \delta_{a b}^{\perp} \delta^{\perp}_{cd},
\end{equation}
where $\delta^{\perp}$ denotes a Kronecker delta only in the perpendicular direction (and zero otherwise). In combination with the prefactor coming from the graviton operator in \eqref{eq:gravop} we therefore have the overall tensor structure
\begin{equation}
T_{\mu \nu \,cd }(\hat{x}) \coloneqq \Big[ \delta_{ac}^{\perp} \delta_{b d}^{\perp} - \frac{1}{2} \delta_{a b}^{\perp} \delta^{\perp}_{cd}\Big]  
\times  \Big[\frac{\sqrt{2}\,\hat{x}^a}{1 + \hat{x}^3} n_{\mu} + \delta^a_{\gap\mu} \Big]\Big[\frac{\sqrt{2}\,\hat{x}^b}{1 + \hat{x}^3}n_{\nu} + \delta^b_{\gap \nu} \Big].
\end{equation}
This is a universal tensorial prefactor to the waveform, which also imposes lightfront gauge on the gravitational radiation. It's now possible to substitute this into the above expression and recognise the overall structure in $k_{\perp}$. Notice that the exponent in the last line of \eqref{eq:Idef} is quadratic in $k_{\perp}$ and therefore the integral is actually a Gaussian of the form 
\begin{multline}
    I_{\mu \nu}(x, y^{-},k_{+}):=-i\int d^2k_{\perp} e^{\im k_a k_b A^{ab}/2k_+ +\im  k_c B^c + \im k_+ C} \\ \times T_{\mu \nu }^{\gap \gap cd}\bigg[k_+ \alpha_{1\, cd} +k_{i}\beta^{i}_{cd}+\frac{k_{i}k_{j}\Gamma^{ij}_{cd}}{k_+} + \alpha_{2 \, cd} \bigg]
\end{multline}
with coefficients defined as
\begin{equation} \label{eq:bigdef}
\begin{gathered}
A^{ij}\coloneqq \int^{+\infty}_{y^-} ds \gamma^{ij}(s)\\
B^{i}\coloneqq\frac{\sqrt{2}(c^{-1})^{i}_{a}\hat{x}^{a}}{1-\hat{x}^{3}} - b^i+\frac{p_{j}}{p_{+}} \int^{y^{-}}ds \gamma^{ij}(s) \\
C \coloneqq \frac{ \sqrt{2} u}{1 - \hat{x}^3} + \frac{ \hat{x} \cdot (c^{-1})b \cdot \hat{x}}{(1 - \hat{x}^3)^2}   - b^+ -\frac{ m^2 y^-}{2 p_+^2}\\ \qquad \qquad \qquad -  \frac{ p_i p_j }{p_+^2} \int^{y^-} \d s\, \gamma^{ij}(s)\\
 \alpha_{1\, cd}\coloneqq E^i_c (y^-)\,  E^j_d (y^-) p_{i}p_{j} \, 
 \\
 \alpha_{2\, cd}\coloneqq- i p_+^2 \sigma_{cd} (y^-) 
 \\
\beta^{i}_{cd}\coloneqq-2 p_{+} \:  E^i_c (y^-)\,  E^j_d (y^-)  p_j   \\
 \Gamma^{ij}_{cd}\coloneqq p_{+}^2  E^i_c (y^-)\,  E^j_d (y^-) .
 \end{gathered}
\end{equation}
Here we have pulled out all $k_+$ dependence from these terms, which will be helpful later when evaluating the $k_+$ integral. Doing the $k_{\perp}$ integral is now an application of Gaussian integration and we obtain 
\begin{equation}
I_{\mu \nu}(k_+) = \mathcal{K} \, e^{-i k_{+}\frac{ BA^{-1}B}{2} + i k_+ C}
\bigg(k_+ \, N_{1\, \mu \nu} +N_{2 \, \mu \nu} \bigg)
\end{equation}
where we have defined the $k_+$ independent quantities
\begin{align}
N_{1\, \mu \nu} &\coloneqq   T_{\mu \nu}^{\gap \gap cd}(\hat{x}) \big(\alpha_{1\, cd} - \beta^i_{cd} A^{-1}_{ij} B^j \nonumber \\& \qquad \qquad+ \Gamma^{ij}_{cd} (A^{-1} B)_i (A^{-1} B)_j\big), \\
N_{2\, \mu \nu} &\coloneqq  T_{\mu \nu}^{\gap \gap cd} (\hat{x})\big(\alpha_{2\, cd} + i \Gamma_{cd}^{ij} A^{-1}_{ij} \big),
\end{align}
and the scalar prefactor 
\begin{equation}
\mathcal{K} =\frac{2 \pi}{\sqrt{\det A}}.
\end{equation}
Note that all of these quantities depend intrinsically on both $x$ and $y$, as well as the memory matrix $c$.
We are now left with the $k_+$ integral, where we can benefit from the simplicity of the $k_+$ dependence in this integrand.

This integral depends on the reality of the integrand. Consider an integral of the form 
\begin{equation}
    2 \Re \int_0^\infty \hat{\d} k_+ \mathcal{N}  e^{-i k_+ \bar{\sigma}(x, y)} \quad , \quad \mathcal{N} \in \mathbb{C} \, ,
\end{equation}
(The $1/k_+$ term can be extracted by integrating with respect to $y$.) When the prefactor $\mathcal{N} \in \mathbb{R}$, this  becomes the usual integral for a $\delta$-function:
\begin{equation}
    \mathcal{N}\int_{-\infty}^\infty \hat{\d} k_+  e^{-\im k_+ \bar{\sigma}(x, y)} =\mathcal{N}  \delta (\bar{\sigma}(x, y)) \quad , \quad \mathcal{N} \in \mathbb{R} \: .
\end{equation}
However, when $\mathcal{N} \in i \mathbb{R}$ , the integrals do not combine, instead giving
\begin{equation}
    \mathcal{N}\Bigg[ \int_{- \infty}^0 \hat{\d} k_+ e^{-\im k_+ \bar{\sigma}(x, y)} - \int_{0}^{\infty} \hat{\d} k_+ e^{-\im k_+ \bar{\sigma}(x, y)} \Bigg]= \mathsf{pv}\Big( \frac{\mathcal{N}}{i \pi \, \bar{\sigma}(x, y)}\Big) \quad , \quad \mathcal{N} \in i \mathbb{R}.
\end{equation}
The $N_1$ term results in a Dirac $\delta$-function (or $\mathsf{pv}(1/\pi\bar{\sigma}(x, y)$), whilst the $N_2$ (also called tail term) can be computed using the prescription
\begin{equation}\label{eq:intrinsic-frame}
\int_{-\infty}^{+\infty} \hat{d k_+} \frac{e^{ \pm i k_+ \bar{\sigma}(x, y)}}{k_+-i \varepsilon}=i  \Theta( \pm \bar{\sigma}(x, y))
\end{equation}
(or $\mathsf{pv} \log (\pm \bar{\sigma}/\pi)$ if the prefactor is imaginary as above).
This prescription around the pole $k_+=0$ is equivalent to a choice of BMS frame often called the ``intrinsic frame" \cite{DiVecchia:2022owy}. 

Firstly, let's assume that $m = 0, 2$. This means that $\sqrt{\det(c)} = e^{i m \pi/2} \sqrt{|\det(c)|}$ is real and therefore taking the real part in \eqref{appheq} we do the integral as for a real\footnote{We additionally assume that $\det(A)>0$.} prefactor $\mathcal{N}$ explained above:
\begin{multline}
h^m_{\mu \nu}(x) \Big|_{\scri^+} = \frac{\sqrt{2} \kappa\, e^{-i m \pi/2}}{2rp_+ \,(1 - \hat{x}^3)\sqrt{|\det(c)|}}\\ \times 
\int_{x_i^-}^{x_f^-} \d y^- \,\mathcal{K} \, \bigg[N_{1\, \mu \nu}\delta\bigg(C - \frac{B A^{-1}B}{2} \bigg)  + i N_{2\, \mu \nu} \Theta \bigg(C - \frac{B A^{-1}B}{2}\bigg)  \bigg] .
\end{multline}
We can now use the fact that the quantity $C - B A^{-1} B/2$ is proportional to Synge's world-function $\sigma(x, y)$ for a particle following geodesic motion and a point on $\scri^+$, as shown in Appendix \ref{app:Synge}. The final result for the waveform when $m = 0, 2$ is then
\begin{multline}
h^m_{\mu \nu}(x)\Big|_{\scri^+} = \frac{\sqrt{2}  \pi \kappa e^{-i m \pi /2}}{r p_+ \, (1 - \hat{x}^3) \, \sqrt{|\det (c A )|}}  \\
\ \times\int_{y_i}^{y_f} \d y^- \bigg[N_{1\, \mu \nu}\delta\big( \bar{\sigma}(x, y^-) \big) + i N_{2\, \mu \nu} \Theta \big( \bar{\sigma}(x, y^-) \big) \bigg],
\end{multline}
where $\bar{\sigma}(x, y^-) = -\sigma(x, y(y^-))/(x^- - y^-)$, the \emph{reduced} Synge's world-function as defined in Appendix \ref{app:Synge}. 

We now repeat this analysis when $m = 1$. Since $\sqrt{\det(c)} = e^{i m \pi/2} \sqrt{|\det(c)|}$ is purely imaginary we consider an imaginary $\mathcal{N}$ and obtain 
\begin{multline}
h^m_{\mu \nu}(x)\Big|_{\scri^+} = \frac{\sqrt{2}  \pi \kappa }{r p_+ \, (1 - \hat{x}^3) \, \sqrt{|\det (c A )|}}  \\
\ \times\int_{y_i}^{y_f} \d y^- \bigg[N_{1\, \mu \nu}\mathsf{pv}\Big(\frac{1}{ \pi \,\bar{\sigma}(x, y^-)} \Big) + i N_{2\, \mu \nu}\mathsf{pv} \log \big( \pi \,\bar{\sigma}(x, y^-)\big) \bigg].
\end{multline}
The main difference to $m =0, 2$ is the functional form of this expression which is consistent with the structure of Green's functions on spacetimes with an odd number of caustics~\cite{Harte:2012uw,Casals:2009zh}.

We now highlight the dependence on memory in this expression. Memory is here captured in the matrix $c_{i a}$, where weak memory is the limit $c_{i a} \rightarrow 0$ with $m =0$. The terms $A$ and $B$, defined in \eqref{eq:bigdef}, are the only quantities with explicit dependence on $c$. However, they are singular as $c \rightarrow 0$, so it is prudent to define the conjugated terms 
\begin{align}
\bar{A}_{a}^{\, \, j}(c) &\coloneqq c_{a i} A^{ij},\label{eq:barA} \\
\bar{B}_a(c) &\coloneqq c_{a i} B^i, \label{eq:barB}
\end{align}
which remain finite in the weak memory limit but still depend on $c$ otherwise. The waveform coefficients $N_{1, 2}$ written in terms of these quantities have explicit dependence on the memory
\begin{align}
N_1 &= T \big(\alpha_1 - \beta^i (\bar{A}^{-1})^{b}_{\, \, i} \bar{B}_b\label{eq:memN1} \\ & \qquad \quad  + \Gamma^{ij} (\bar{A}^{-1} \bar{B})_i (\bar{A}^{-1} \bar{B})_j\big), \nonumber \\
N_2 &=T \big( \alpha_2 + \im \, \Gamma^{ij} c_{ai}(\bar{A}^{-1})^{a}_{\, \, j}\big), \label{eq:memN2}
\end{align}
neglecting the indices on $T_{\mu \nu}^{\gap \gap cd}$ and other quantities to save space. With these expressions in place, taking the weak memory limits just corresponds to switching off the higher orders in $c$, which leaves the no-memory expression~\cite{Adamo:2022qci} as a special case
\begin{multline}
h_{\mu \nu}(x) \Big|_{\scri^+} = \frac{ \sqrt{2} \pi \kappa}{r \, (1 - \hat{x}^3)\,p_{+} \det(\bar{A})} \\
\times   \, \int_{y_i}^{y_f} \d y^- \Bigg[N_{1\, \mu \nu} \delta(\bar{\sigma}(x, y^-)) + \im T_{\mu \nu}^{\gap \gap ab} (\hat{x}) \alpha_{2 \, ab} \Theta(\bar{\sigma}(x, y^-))\Bigg].
\end{multline} 
\section{Synge's world function in plane wave spacetimes} \label{app:Synge}
Synge's world function $\sigma(x, x') = \sigma(x', x)$ is a key object in the construction of Green's functions in any space-time. It encodes the geodesic distance between two space-time points $x, x'$, and in plane waves can be shown~\cite{Harte:2015ila} to take the form in mostly negative signature
\begin{equation}
\frac{\sigma(x, y)}{x^- - y^-} \coloneqq x^+ + \frac{\sigma_{ab}(x^-)}{2}x^a x^b - y^+ - \frac{\sigma_{ab}(y^{-})}{2} y^a y^b -  \frac{1}{2}\mathcal{B}^i (x, y) (\mathcal{A}^{-1}(x, y))^{ij} \mathcal{B}^j(x, y),
\end{equation}
where we define
\begin{gather}
\mathcal{B}^i(x, y) \coloneqq E^i_a(x^-)x^a - E^i_a(y^-)y^a, \\
\mathcal{A}^{ij} \coloneqq \int^{x^-}_{y^-} \gamma^{ij}(s) \d s
\end{gather}
and $\sigma(x, y)$ can be shown to be independent of the boundary conditions for the geometric objects used in these definition.

In this work, we require the behaviour of $\sigma(x, y)$ as one of the points $y$ approaches $\scri^+$ and the other follows massive geodesic motion on the background. The former requires us to take the asymptotic limit of all quantities dependent on $x$ in this expressions according to the Bondi coordinates in Appendix \ref{AppBondi}. The latter condition requires us to evaluate $y$ on the path of the particle determined by its initial momentum and parametrised by $y^-$. This is most easily solved first in Einstein-Rosen coordinates (see~\cite{Harte:2015ila}) and then converted to Brinkmann coordinates. The geodesic motion is written implicitly as
\begin{gather}
y^+(y^-) + \frac{\sigma_{ab}(y^-)}{2} y^a y^b = b^+ \nonumber + \frac{m^2 y^-}{2p^{2}_+} \quad \quad  \\\qquad \qquad + \frac{p_i p_j}{2p_+^{2}} \int^{y^-} \d s \, \gamma^{ij}(s)\\
E^j_a (y^-) y^a(y^-) = b^j -  \frac{p_i}{p_+}\int^{y^-} \d s \, \gamma^{ij}(s)\,.
\end{gather}
where $b$ encodes the relative position of the trajectory to the coordinate system at initial times. For the $x$-coordinates, taking $r \rightarrow \infty$ and representing $x$ with the boundary coordinates $u, \hat{x}$ we can use the large distance expressions found in Appendix \ref{AppBondi} to write
\begin{multline} \label{eq:appRedSynge}
- \frac{ \sigma(x, y)}{x^- - y^-}\Bigg|_{x \in \scri^+} = \frac{ \sqrt{2} u}{1 - \hat{x}^3} + \frac{ \hat{x} \cdot (c^{-1})b \cdot \hat{x}}{(1 - \hat{x}^3)^2}- b^+ - \frac{m^2 y^-}{2p_+^2} \\- \frac{p_i p_j}{p_+^2} \int^{y^-} \d s \gamma^{ij}(s)  - \frac{1}{2}B^i (A^{-1})_{ij} B^j,
\end{multline}
where
\begin{gather}
B^i = \frac{\sqrt{2} (c^{-1})^i_a \hat{x}^a}{1 - \hat{x}^3} - b^i + \frac{p_j}{p_+} \int^{y^-} \d s \, \gamma^{ij}(s), \\
A^{ij} = \int^{\infty}_{y^-} \d s \,\gamma^{ij}(s).
\end{gather}
This function represents the geodesic distance between a massive particle following geodesic motion with initial momentum $p$, evaluated at light-front time $y^-$, and a point on $\scri^+$ with coordinates $(u, \hat{x})$, as desired.

\bibliography{refs}
\bibliographystyle{JHEP}
\end{document}